\documentclass[aps,prd,twocolumn,groupedaddress,showpacs,showkeys]{revtex4}

\usepackage{graphicx}
\usepackage{psfrag}
\usepackage{subfigure}
\usepackage{amsfonts}
\usepackage{latexsym}

\begin{document}


\title{Interacting Dipoles from Matrix Formulation of Noncommutative Gauge Theories}

\author{Li~Jiang}
\email{jiang@physics.utexas.edu}
\author{Eric~Nicholson}
\email{ean@physics.utexas.edu} \affiliation{Theory Group, Department of Physics,
             University of Texas at Austin,
             TX 78712, USA}

\date{\today}

\begin{abstract}
We study the IR behavior of noncommutative gauge theory in the matrix formulation. In this approach, the nature of
the UV/IR mixing is easily understood, which allows us to perform a reliable calculation of the Wilsonian quantum
effective action for the noncommutative gauge field. We find non-trivial interaction terms suggestive of the
dipole degrees of freedom that are expected from the decoupling limit of open string theory in a strong NS-NS
$B$-field. Furthermore, the leading IR interactions are mediated by virtual UV states, which sheds new light on
UV/IR mixing and the non-analytic dependence of the quantum theory on the noncommutativity parameter, $\theta$.
\end{abstract}



\maketitle

\section{Introduction}
Noncommutative quantum field theories have been extensively studied in the past few years.  The motivation to study these
systems stems from the fact that noncommutative gauge theories arise naturally from string theory through various
decoupling limits \cite{sw:stri}\cite{sb:magn}. However, the infrared behavior of noncommutative quantum field theories
remains poorly understood due to a UV/IR connection in which the IR dynamics is not decoupled from the UV.  In particular,
the UV region of loop integration in Feynman diagrams leads to non-analytic behavior in external momenta indicative of novel IR
dynamics \cite{ms:nonp}\cite{ls:gaug}.

Recently, the insight of \cite{ik:bi} and \cite{mv:mean} as well as \cite{ki:inte}\cite{ar:uvir} has shed some
light on the interpretation of the leading IR singularities that occur in non-su\-per\-sym\-met\-ric
non\-com\-mut\-a\-tive theories. The authors of \cite{ik:bi}\cite{mv:mean}, working in the matrix formulation of
noncommutative gauge theory, were able to match the leading one-loop IR singularity with an instantaneous two-body
interaction between gauge invariant operators. Moreover, their results had a natural interpretation in terms of
string and brane degrees of freedom that appear in matrix theory.

Actually, in the decoupling limit of open string theory in which NCYM emerges as the low energy effective theory,
the remaining degrees of freedom are known to be extended objects \cite{sb:magn}. In particular, the quanta can be
thought of as dipoles with a transverse size proportional to their center of mass momentum. In fact, this is the
origin of UV/IR mixing: high momentum dipoles grow long in spatial extent. We, therefore, intuitively expect
instantaneous interactions between distant points mediated by long dipoles. However, in the conventional star
product approach to noncommutative field theory, the intrinsic dipole structure of the elementary quanta is far
from clear, although some suggestive results have been obtained \cite{ki:inte}.

Perhaps the most important lesson gleaned from \cite{ik:bi}\cite{mv:mean} is that the matrix formulation is the
most natural framework in which to study noncommutative gauge theory. While \cite{mv:mean} did not demonstrate the
intrinsic dipole structure of NCYM in the sense of \cite{sb:magn}, it did show that the noncommutative gauge
invariance is manifest, which leads to immense simplification at the technical level. Moreover, since matrix
models naturally describe extended objects, one might hope that the intrinsic dipole structure could be made
manifest as well, similar to the spirit of the bi-local representation discussed in \cite{ik:bi}. Surely, this
would lead to great conceptual clarity regarding the physics of NCYM. With this in mind, we seek to study
noncommutative gauge theory in the matrix formulation in order to develop a better intuition for the physics of
dipole theories and the corresponding UV/IR connection.

However, there are some technical as well as conceptual obstacles to be overcome. At the technical level, problems
typically arise because conventional field theory techniques often lead to ambiguous IR behavior, essentially
because, in the light of noncommutativity, UV and IR are no longer synonymous with short distance and long
distance, respectively. Therefore, in order to proceed, we will have to develop new calculational tools which will
allow us to calculate, in a straight forward fashion, terms in the quantum effective action. We are then left to
interpret the results. The conceptual challenge is then to understand the matrix calculation in conventional field
theory terms, in addition to identifying the effects of the fundamental dipole structure and the corresponding
UV/IR mixing.

In order to develop an intuition for the IR behavior of noncommutative gauge theory, we calculate the Wilsonian
quantum effective action for the gauge field, in the matrix approach. After deriving the matrix propagator, we
proceed with perturbative calculations, which yield interaction terms suggestive of dipole degrees of freedom. As
expected, these dipoles have a length proportional to their center of mass momentum, and therefore, integrating
out UV dipoles will lead to instantaneous long distance interactions. In fact, the leading long distance
interactions, that dominate in the IR, will be due to the virtual UV dipoles. We are finally left with a very
clear and intuitive picture of the dynamics resulting from dipole degrees of freedom and UV/IR mixing, which is
reminiscent of the bi-local field representation discussed in \cite{ik:bi}.

This paper will be organized as follows. In section 2, we review the matrix formulation of noncommutative gauge
theories. In section 3, we review the background field gauge fixing that we employ in later sections in order to
calculate the quantum effective action.  In section 4, we derive the propagator in the matrix quantum mechanics
and then recast it in a new set of variables more suitable for a field theory interpretation. In section 5, we
calculate the one-loop quantum effective action and interpret the interactions in terms of dipole degrees of
freedom. In section 6, we discuss higher loop quantum corrections as well as the difference between our dipole
interpretation and that of \cite{ki:inte}.  We end with some discussion and outlook. The appendices contain some
technical details.

\section{Matrix formulation of NCYM}

Consider the following Lagrangian describing a form of $U(M)$ symmetric matrix quantum mechanics
\begin{eqnarray} 
\lefteqn{ L={\rm Tr}\left(\frac{1}{2}(\dot{X^{i}}-i[A_0,X^{i}])^{2} \right. } \nonumber\\
 & & \left. \qquad\qquad{}+\frac{1}{4}[X^{i},X^{j}][X^{i},X^{j}]+\ldots \right)
\end{eqnarray}
where $(A_0,X^{i})$ are $M\times M$ hermitian matrices transforming in the vector representation of $SO(2p,1)$ and the
adjoint representation of $U(M)$.  The $\ldots$ represent other fields such as the fermions in the supersymmetric theory.
However, in the following we will discuss only the treatment of the bosonic fields, the generalization to fermions
being obvious.

In the $M\rightarrow\infty$ limit, we can consider the classical ground state given
by $X^{i}=\hat{x}^{i}\otimes {\rlap{1} \hskip 1.6pt \hbox{1}}_{N\times N}$ and $A_0=0$ where $\hat{x}^{i}$ are
time-independent hermitian matrices satisfying the algebra of the noncommuting $2p$-plane
\begin{equation}
[\hat{x}^{i},\hat{x}^{j}]=i{\theta}^{ij}{\rlap{1} \hskip 1.6pt \hbox{1}}.
\end{equation}
${\theta}^{ij}$ is a real constant anti-symmetric tensor of $SO(2p)$. Following \cite{ns:back}, we expand in
fluctuations about this background $X^{i}=\hat{x}^{i}\otimes {\rlap{1} \hskip 1.6pt \hbox{1}}_{N\times N}+
{\theta}^{ij}A_j(\hat{x})$ and $A_0=A_0(\hat{x})$.  The resulting action describes NCYM
\begin{eqnarray}
\lefteqn{ L=\int d^{2p}x\,{\rm tr}_N\left(\frac{1}{2}G^{ij}F_{0i}(x)F_{0j}(x) \right.} \\
 & & \left.{}-\frac{1}{4}G^{ij}G^{kl}(F_{ik}(x)-{{\theta}^{-1}}_{ik})(F_{jl}(x)-{{\theta}^{-1}}_{jl})+\ldots\right) \nonumber
\end{eqnarray}
where $F_{\mu\nu}(x)=\partial_{\mu}A_{\nu}(x)-\partial_{\nu}A_{\mu}(x)-iA_{\mu}(x)*A_{\nu}(x)+iA_{\nu}(x)*A_{\mu}(x)$ is
the noncommutative field strength, and $G^{ij}={\theta}^{ik}{\theta}^{kj}$ is the inverse spatial metric. In deriving
the NCYM theory, we have used the standard map between ordinary  coordinates and noncommuting matrix
coordinates \cite{rg:soli}
\begin{eqnarray}
A(x)*B(x) & \longleftrightarrow &  A(\hat{x})B(\hat{x}); \nonumber\\
\int d^{2p}x\,{\rm tr}_N\left(A(x)*B(x)\right) & \longleftrightarrow & {\rm Tr}\left(A(\hat{x})B(\hat{x})\right).
\end{eqnarray}
Note that, for notational simplicity, we have set $(2\pi)^{2p}\det(\theta)=1$.

Thus, $2p+1$ dimensional NCYM with a constant background field strength can be described by $0+1$ dimensional matrix
quantum mechanics.  From this point on, we will work almost exclusively in the matrix picture; however, we will
eventually arrive at an interpretation of the dynamics in $2p+1$ dimensions.

\section{Background field gauge fixing of the matrix model}

Ultimately, we will be interested in the IR behavior of the noncommutative gauge field in the quantum theory. We
can systematically compute the quantum effective action by expanding the fields $A_0=B_0+A$ and
$X^{i}=B^{i}+Y^{i}$ where $B_0$ and $B^{i}$ are background fields satisfying the equations of motion, while $A$
and $Y^{i}$ are fluctuating fields to be integrated out. For our purpose, we will specialize to backgrounds of the
form $B_0=0$ and $B^{i}={\hat{x}}^{i}\otimes {\rlap{1} \hskip 1.6pt \hbox{1}}_{N\times
N}+{\theta}^{ij}A_{j}(\hat{x})$.

In order to define the functional integral over $A$ and $Y^{i}$, we must gauge fix the Lagrangian. This can be
accomplished by adding both a gauge fixing and a ghost term to the action
{\setlength\arraycolsep{2pt}
\begin{eqnarray}
L_{\mathrm{gf}} & = & {\rm Tr}\left(-\frac{1}{2}(-\dot{A}-i[B^{i},Y^{i}])^{2}\right);\nonumber\\
L_{\mathrm{gh}} & = & {\rm Tr}\left(\dot{\bar{c}}(\dot{c}-i[A,c])+[B^{i},\bar{c}][X^{i},c]\right).
\end{eqnarray}}\\
Upon expanding in fluctuations, the action takes the form $L=L_0+L_2+L_3+L_4$ where
{\setlength\arraycolsep{2pt}
\begin{eqnarray}
L_0 & = & {\rm
Tr}\left(\frac{1}{2}{\dot{B}}^{i2}+\frac{1}{4}[B^{i},B^{j}][B^{i},B^{j}]\right);
\nonumber\\
L_2 & = & {\rm Tr}\left(\frac{1}{2}{\dot{Y}}^{j2}+\frac{1}{2}[B^{i},Y^{j}]^{2}-\frac{1}{2}{\dot{A}}^{2}-\frac{1}{2}[B^{i},A]^{2}+\dot{\bar{c}}\dot{c} \right. \nonumber\\
 & & \left.{}+[B^{i},\bar{c}][B^{i},c]+[B^{i},B^{j}][Y^{i},Y^{j}]-2i{\dot{B}}^{i}[A,Y^{i}] \right); \nonumber\\
L_3 & = & {\rm Tr}\left([B^{i},A][A,Y^{i}]+[B^{i},Y^{j}][Y^{i},Y^{j}]+[B^{i},\bar{c}][Y^{i},c] \right. \nonumber\\
 & & \left.{}-i{\dot{Y}}^{i}[A,Y^{i}]-i\dot{\bar{c}}[A,c] \right); \nonumber\\
L_4 & = & {\rm Tr}\left(\frac{1}{4}[Y^{i},Y^{j}][Y^{i},Y^{j}]-\frac{1}{2}[A,Y^{i}]^{2}\right).
\end{eqnarray}}\\
From $L_2$, we see that all of the fluctuating fields have similar quadratic terms up to terms proportional
to $i{\dot{B}}^{i}$ and $[B^{i},B^{j}]$. If we write the background field in terms of the noncommutative gauge
field, $B^{i}={\hat{x}}^{i}\otimes {\rlap{1} \hskip 1.6pt \hbox{1}}_{N\times N}+{\theta}^{ij}A_j(\hat{x})$, we find that
\begin{equation}\label{eq:field}
i{\dot{B}}^{i}=i{\theta}^{ij}F_{0j}\qquad\, [B^{i},B^{j}]=i{\theta}^{ik}{\theta}^{lj}(F_{kl}-{{\theta}^{-1}}_{kl}).
\end{equation}
Thus, these terms are proportional to the background gauge field strength. As is well known, the background field
dependence of the terms quadratic in the fluctuating fields can either be treated exactly or perturbatively,
depending on the definition of the propagator. In our calculation, it will be most convenient to treat the field
strength terms (\ref{eq:field}) perturbatively, while absorbing the remaining background dependence into the
propagator. From a physical standpoint, this choice corresponds to a derivative expansion of the background field.
However, we will consider only the leading order long distance interactions, in which case commutator terms will
be suppressed.

\section{Matrix propagator}

As discussed above, all of the fluctuating fields, $\Phi=(Y^{i},A,\bar{c},c)$ corresponding to gauge field degrees
of freedom, have similar quadratic terms of the form
\begin{equation}
{\rm Tr}\left(\frac{1}{2}{\dot{\Phi}}^{2}+\frac{1}{2}[B^{i},\Phi]^{2}\right).
\end{equation}
In terms of indices living in the fundamental and anti-fundamental representations of $U(\infty)$, the adjoint
matrix $\Phi={{\Phi}_{b}}^{a}$.  In matrix notation, the quadratic term becomes
{\setlength\arraycolsep{2pt}
\begin{eqnarray}
 & & \frac{1}{2}{{\Phi}_{b}}^{a}\left(-{{\delta}_{c}}^{b}{{\delta}_{a}}^{d}\frac{d^{2}}{{dt}^{2}}-{{B^{i}}_{e}}^{b}{{B^{i}}_{c}}^{e}{{\delta}_{a}}^{d}
 \right. \nonumber\\
 & & \qquad \qquad \left. -{{\delta}_{c}}^{b}{{B^{i}}_{e}}^{d}{{B^{i}}_{a}}^{e}+2{{B^{i}}_{c}}^{b}{{B^{i}}_{a}}^{d}\right){{\Phi}_{d}}^{c} \nonumber\\
 & = & \frac{1}{2}{\Phi}^{T}\left(-{\rlap{1} \hskip 1.6pt \hbox{1}}\otimes {\rlap{1} \hskip 1.6pt \hbox{1}} \frac{d^{2}}{{dt}^{2}}-(B^{i}\otimes {\rlap{1} \hskip 1.6pt \hbox{1}}-{\rlap{1} \hskip 1.6pt \hbox{1}}\otimes B^{i})^{2}\right)
\Phi.
\end{eqnarray}}\\
In the Wilsonian scheme, we are only  interested in integrating out virtual states with frequencies $\omega\gg
1/T$, $T$ being the time scale set by the background. For these high frequency modes, the back-reaction coming
from the background time dependence is a subleading effect. Therefore, the matrix propagator for virtual states
with frequencies above a Wilsonian cutoff, $\Lambda\gg 1/T$, can be expressed in the following Fourier integral
form
\begin{equation}
G(t-t^{\prime})=\int_{\Lambda}\frac{d\omega}{2\pi}\frac{e^{-i\omega (t-t^{\prime})}}{{\omega}^{2}-M^{2}}+\ldots
\end{equation}
where  $M^{2}=(B^{i}\otimes {\rlap{1} \hskip 1.6pt \hbox{1}}-{\rlap{1} \hskip 1.6pt \hbox{1}}\otimes B^{i})^{2}$ and
the $\ldots$ represent subleading terms that are suppressed by factors of $(T\Lambda)^{-1}\ll 1$.  In the following,
we consider only the leading order term.

To relate this $0+1$ dimensional matrix quantity to a $2p+1$ dimensional field theory quantity, we choose a
convenient representation which is derived in
APPENDIX \ref{ap:propagator}
{\setlength\arraycolsep{2pt}
\begin{eqnarray}
& & \frac{1}{{\omega}^{2}-M^{2}} \nonumber\\
& = & \int\frac{d^{2p}k}{(2\pi)^{2p}}e^{-ik\cdot(B\otimes 1-1\otimes B)}
\int_{\theta\Lambda}d^{2p}x \frac{e^{ik\cdot x}}{{\omega}^{2}-x^{2}}\\
& = & \int\frac{d^{2p}k}{(2\pi)^{2p}}e^{-ik\cdot B}\otimes e^{ik\cdot B}\int_{\theta\Lambda} d^{2p}x\,e^{ik\cdot
x}\widetilde{G}(\omega,{\theta}^{-1}x)\nonumber
\end{eqnarray}}\\
where $\widetilde{G}(\omega,p)=({\omega}^{2}-p_i G^{ij} p_j)^{-1}$ is the field theory momentum space propagator
for a massless state.  As discussed in the appendix, there is a lower cutoff applied to the integral over $x$ such
that $x>\theta\Lambda\gg{\theta}/{L}$ where $L$ is the length scale set by the background. Putting everything together,
the matrix propagator can be written in the
following form
\begin{eqnarray} \label{eq:prop}
\lefteqn{ G(t-t^{\prime})=\int_{\theta\Lambda} d^{2p}x \int_{\Lambda}\frac{d\omega}{2\pi}\int\frac{d^{2p}k}{(2\pi)^{2p}} e^{-i\omega(t-t^{\prime})+
ik\cdot x} } \nonumber\\
 & & \qquad\qquad\qquad {}\times \widetilde{G}(\omega,{\theta}^{-1}x)e^{-ik\cdot B}\otimes e^{ik\cdot B}.\qquad
\end{eqnarray}
To our knowledge, neither this representation of the propagator, nor the interpretation to follow has been
previously recognized. However, our approach is reminiscent of the work in \cite{ik:bi} regarding bi-local fields.

We can now identify the various ingredients of the $0+1$ dimensional propagator from a $2p+1$ dimensional
perspective. As suggested by the notation, $(\omega,k)$ is to be identified with the spacetime energy momentum;
likewise, $(t,x)$ is the corresponding spacetime coordinate. The integral over $x$ is then understood in terms of
the nonlocality of the noncommutative field theory.  Perhaps more surprising is the role played by the field
theory propagator, $\widetilde{G}$. Evidently, the small $k$/large $x$ region of the integral corresponding to low
momentum/large distance receives contributions from \emph{high} momentum field theory states and vice-versa.
Actually, this type of behavior has a very natural interpretation in terms of the dipole degrees of freedom that
we expect from the decoupling limit of open string theory in a strong NS-NS $B$-field \cite{sb:magn}.

In the decoupling limit, the noncommutative field quanta can be thought of as dipoles with a transverse size
proportional to the center of mass momentum ${x}^{i}={\theta}^{ij}p_j$. It is clear that this effect is encoded in
(\ref{eq:prop}) above, since the momentum argument of the field theory propagator, $\widetilde{G}(\omega,p)$, is
$p={\theta}^{-1} x$. It is also clear that, due to the Fourier integral over position, these dipole states probe a
transverse momentum scale $k_{i}\sim 1/{{x}^{i}}$. Combining these two relations, we arrive at $1\sim
p_{i}{\theta}^{ij}k_{j}$, which is a familiar result from the star product formulation \cite{ms:nonp}. In essence,
this relation means that integrating out high momentum states can lead to low momentum effects, which will become
more concrete in subsequent sections. Thus, it seems that (\ref{eq:prop}) naturally describes the dipole degrees
of freedom that appear in NCYM.

However, it is important to realize that this representation is only valid for dipoles of high energy and
momentum. More precisely, if the background changes on time and length scales $T$ and $L$, respectively, we can
only integrate over frequencies $\omega \gg  1/T$ and momenta $p={\theta}^{-1}x \gg 1/L$. Otherwise, the time
derivatives and commutators involving the background field that were dropped in the derivation of (\ref{eq:prop})
are no longer negligible. Therefore, the cutoff $\Lambda$ is chosen such that $\Lambda\gg 1/T \,\textrm{and}\,
1/L$. In this case, the higher order commutator and time derivative corrections are suppressed by factors of
$(L\Lambda)^{-1}\,\,\textrm{and}\,\, (T\Lambda)^{-1}\ll 1$. Moreover, since the cutoff is chosen relative to the
scale of the background, $\Lambda$ is naturally interpreted in the Wilsonian sense.

The matrix structure of the $0+1$ dimensional propagator, which is contained entirely in the tensor product of
operators of the form $\exp(ik\cdot B)$, also has an important field theory interpretation. Using the standard
dictionary between noncommuting matrix coordinates and ordinary coordinates, we can identify \cite{mv:mean}
\begin{eqnarray}
\lefteqn{ e^{ik\cdot B}=e^{ik\cdot\hat{x}\otimes 1_{N\times N}+ik\cdot\theta\cdot A(\hat{x})}\longleftrightarrow }
                    \nonumber\\
 & & \qquad\qquad P_{*}e^{i\int_0^{1}d\sigma k\cdot\theta\cdot A(x+\sigma k\cdot\theta)}*e^{ik\cdot x}
\end{eqnarray}
where $P_{*}$ denotes path ordering of the exponential using the $*$ product.  This object transforms in the
adjoint under gauge transformation, and in particular, the trace is gauge invariant
\begin{equation}\label{eq:rho1}
{\rm Tr}\left(e^{ik\cdot B}\right)\longleftrightarrow \int d^{2p}x e^{ik\cdot x} {\rm
tr}_N\left(P_{*}e^{i\int_0^{1}d\sigma k\cdot\theta\cdot A(x+\sigma k\cdot\theta)}\right).
\end{equation}\\
We immediately recognize this object as an open Wilson line . In fact, this structure was essentially guaranteed
by the noncommutative gauge invariance \cite{ni:obse}\cite{hl:trek}. In later sections, when we use the matrix
propagator in perturbative calculations of the effective action, we will frequently encounter the Fourier
transform of the open Wilson line above. Following \cite{mv:mean}, we define the operator
\begin{equation}\label{eq:rho2}
\rho(x)=\int\frac{d^{2p}k}{(2\pi)^{2p}}e^{ik\cdot x}{\rm Tr}\left(e^{-ik\cdot B}\right).
\end{equation}
Note that although $\rho(x)$ is generally a nonlocal field theory operator, for $\theta\cdot k$ sufficiently small
such that (\ref{eq:prop}) is valid, it is approximately local on length scales given by the background
configuration, as can be easily seen from (\ref{eq:rho1}) and (\ref{eq:rho2}). In fact, all gauge invariant Wilson
line operators, which differ only by extra operator insertions, will share this property.

The interpretation of the matrix propagator in terms of dipole degrees of freedom is made more concrete in the
following section by calculating the Wilsonian quantum effective action. We will find that integrating out UV
virtual states gives rise to long distance interaction terms which are naturally interpreted in the dipole context
discussed above. We will also identify terms that correspond to traditional renormalization of coupling constants
of the theory.

\section{One-loop effective action}

\begin{figure}
\subfigure[Single matrix diagram is manifestly gauge invariant and implicitly contains all leading background
dependence.]{\label{fig:subfig:a}\begin{minipage}[b]{.5\textwidth}\centering
\includegraphics[scale=0.25]{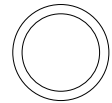}\end{minipage}}

\subfigure[Gauge invariance achieved by summing over all background insertions on both the outer and inner
boundaries.]{\label{fig:subfig:b}\begin{minipage}[b]{.5\textwidth}\centering
\includegraphics[scale=0.25]{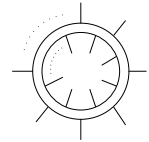}\end{minipage}}
\caption{One-loop contributions in the matrix versus the star product approach.}
\end{figure}

We begin the computation of the quantum effective action at one-loop, where the advantage of the matrix
formulation becomes immediately clear. The leading one-loop contribution is manifestly gauge invariant and can be
expressed in a single diagram drawn in 't~Hooft double line notation as shown in FIG.~\ref{fig:subfig:a}. This is
to be contrasted with the field theory star product approach in which an infinite number of diagrams of the form
shown in FIG.~\ref{fig:subfig:b} must be summed up in order to achieve gauge invariance
\cite{ki:inte}\cite{ar:uvir}\cite{hl:trek}.

Using our representation of the propagator (\ref{eq:prop}), the evaluation of the matrix diagram is simple. The
contraction of matrix indices, as indicated by the double line diagram, gives a double trace contribution
proportional to
{\setlength\arraycolsep{2pt}
\begin{eqnarray}\label{eq:trac}
 & & \int\frac{d\omega}{2\pi}{\rm Tr}\log{G(\omega)} \\
 & = & \int d^{2p}x_1 d^{2p}x_2 \rho(x_1,t)\rho(x_2,t)\int\frac{d\omega}{2\pi}\log{\widetilde{G}(\omega,{\theta}^{-1}x_{12})}
                   \nonumber
\end{eqnarray}}\\
where $x_{12}=x_1-x_2$. Moreover, the contraction of spacetime vector and spinor indices contributes a factor
proportional to $N_B-N_F$ where $N_B$ and $N_F$ are the numbers of on shell bosonic and fermionic polarization
states. Note that, although we have not discussed fermions up to now, the matrix propagator for fermionic fields
can be constructed in the exact same way as for the bosonic fields. Furthermore, the integrals are always assumed
to be cutoff as previously discussed.

Now let us understand the structure of (\ref{eq:trac}) a bit more in terms of conventional field theory diagrams
FIG. \ref{fig:subfig:b}. First of all, we choose to expand $\rho(x)={\rm tr}_{N}({\rlap{1} \hskip 1.6pt
\hbox{1}})+\Delta(x)$. The significance of $\Delta(x)$ is that it contains only fluctuations around the constant
background. In particular, $\Delta(x)$ vanishes for trivial configurations gauge equivalent to $A_{i}(x)=0$, which
can be seen easily from the formulas (\ref{eq:rho1}) and (\ref{eq:rho2}). Therefore, the field theory
interpretation of $\Delta(x)$ is that it represents the gauge invariant contribution from the insertions of the
background gauge field into either the outer or inner boundary of the loop. On the other hand, the constant term
of $\rho(x)$ is gauge field independent, and therefore, must descend from field theory diagrams with no background
insertions on the corresponding outer or inner boundary.

For example, we can conclude that the ${\Delta}^{0}$ interaction involves
no insertions on either the outer or the inner boundary, and therefore, comes from field theory vacuum diagrams.
Using the same reasoning, we find that the ${\Delta}^{1}$ interactions involve background insertions on only one
boundary, and therefore, are due to planar field theory diagrams. Finally, the ${\Delta}^{2}$ interaction involves
insertions on both the outer and the inner boundary, and therefore, arises from non-planar field theory diagrams.
Thus, the single matrix diagram in FIG. \ref{fig:subfig:a} contains contributions from both planar and non-planar
field theory diagrams.

However, the matrix calculation (\ref{eq:trac}) only reproduces the leading order terms of the expansion in
external momenta, as can be verified by a direct field theory calculation \cite{mv:mean}. The reason is that in
deriving the propagator (\ref{eq:prop}), the matrix formulation naturally leads to an expansion in commutators and
time derivatives. The subleading terms, as we have seen,  are suppressed by factors of $(L\Lambda)^{-1}$, $1/L$
being the scale of the external momenta and $\Lambda$ the scale of the Wilsonian cutoff. Clearly, this corresponds
to expanding the field theory diagrams in the external momenta since the expansion parameter is the same. Thus, as
alluded to earlier, the physical nature of our approximation is that of a derivative expansion of the background.
In fact, order by order, the matrix approach reproduces the momentum expansion of the field theory if higher order
commutators and time derivatives are retained. However, the matrix approach is best equipped to describe long
distance behavior, in which case the $(L\Lambda)^{-1}$ corrections are small and the leading order term dominates.

Back to the calculation at hand, as expected, the ${\Delta}^{0}$ and ${\Delta}^{1}$ interactions,  corresponding to
planar field theory diagrams, are divergent. It is easy to see that they are proportional to
\begin{equation}
\int\frac{d\omega}{2\pi} d^{2p}x_{12} \log{\widetilde{G}(\omega,{\theta}^{-1}x_{12})}= \int\frac{d\omega
d^{2p}p}{2\pi(2\pi)^{2p}}\log{\widetilde{G}(\omega,p)}.
\end{equation}
In fact, this is nothing but the usual leading UV divergence that is familiar from field theory. It is important
to note, however, that these divergent terms do not contribute to the dynamics of the background gauge field. The
reason is that $\int d^{2p}x\Delta(x)=0$, as can be seen from (\ref{eq:rho2}). The same argument is given in
\cite{mv:mean} from a different point of view. In any case, these contributions are independent of the background
configuration, so we ignore them.

\begin{figure}
\psfrag{x}{$x_1$}
\psfrag{y}{$x_2$}
\includegraphics[height=1.25cm]{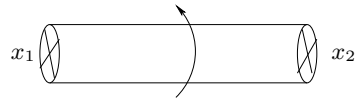}
\caption{High momentum virtual dipoles grow long in the transverse direction and mediate instantaneous interactions between
distant background fluctuations at $x_1$ and $x_2$.} \label{fig:cyli}
\end{figure}

On the other hand, the non-planar diagrams represented by the ${\Delta}^{2}$ interaction,
\begin{equation}\label{eq:non1}
\int d^{2p}x_1 d^{2p}x_2 \Delta(x_1,t)\Delta(x_2,t)\int\frac{d\omega}{2\pi}\log{\widetilde{G}(\omega,{\theta}^{-1}x_{12})},
\end{equation}
are more interesting. This term illustrates how UV dipoles can mediate long distance interactions: when the
virtual dipoles in the loop have high momentum, FIG. \ref{fig:subfig:a} ``stretches out'' into a long cylinder
that joins distant points $x_1$ and $x_2$. Each boundary of the cylinder contributes a trace which yields a gauge
invariant Wilson line operator corresponding to the low momentum background insertions of the field theory
diagrams. This process is depicted in FIG. \ref{fig:cyli}. Thus, we can interpret the double lines in the matrix
diagram as representing the physical separation of the two ends of the dipole quanta.

At this point, however, there is a technicality to be addressed. Since the leading contribution to the effective
potential between $\Delta(x_1)$ and $\Delta(x_2)$ grows strong at large separation
\begin{equation}
\int \frac{d\omega}{2\pi}\log\widetilde{G}(\omega,{\theta}^{-1}x_{12})\sim|x_1-x_2|+\textrm{constant},
\end{equation}
the theory, in the presence of this term, is strongly interacting at long distances.  This fact has been
recognized in \cite{mv:mean}, and it was shown that these strong long distance interactions are due to the leading
IR pole singularities that appear in non-supersymmetric noncommutative theories. The significance of the poles has
also been discussed in \cite{ki:inte}\cite{ar:uvir}. However, we seek a weakly coupled long distance description,
since otherwise, we can not treat the system perturbatively. Therefore, we demand $N_B=N_F$, in which case the
leading interaction cancels.

We must now consider the next to leading order one-loop contribution, which has also been discussed from the star
product perspective in \cite{ar:uvir}. As alluded to earlier, the precise result requires that we keep the next to
leading order commutators that were dropped in the derivation of the propagator, as well as extra insertions of
the background field strength coming from terms in $L_2$ that were also excluded from the propagator. However,
power counting as well as symmetry arguments imply that the next to leading order one-loop contribution will be of
the same order as FIG. \ref{fig:subfig:a} with two extra insertions of the field strength
\begin{center}
\includegraphics[scale=0.25]{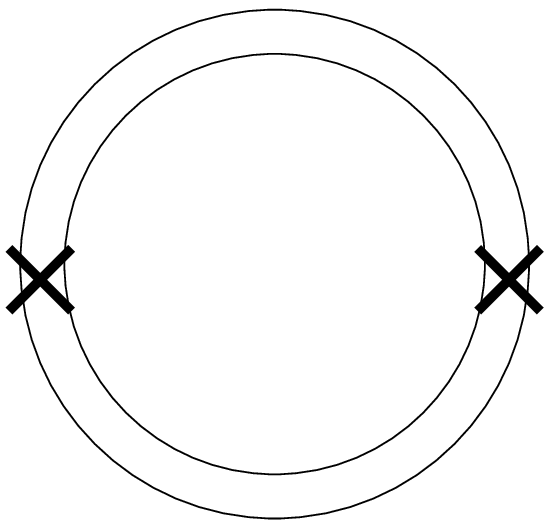}.
\end{center}
This contribution alone suffices to demonstrate the qualitative features of the next to leading order one-loop
behavior. It has the added virtue that the calculation can be done with the propagator (\ref{eq:prop}) because the
field strength insertions are already higher order. A straight forward calculation outlined in APPENDIX
\ref{ap:one-loop} gives a term in the action proportional to
\begin{eqnarray}\label{eq:ntlo}
\lefteqn{\int dt d^{2p}x_1 d^{2p}x_2\left[{\rho}_{FF}(x_1,t)\rho(x_2,t)-{\rho}_{F}(x_1,t){\rho}_{F}(x_2,t)\right]}
                                                 \nonumber\\
 & & \qquad\quad \times\int\frac{d\omega}{2\pi}
\widetilde{G}({\omega},{\theta}^{-1}x_{12})\widetilde{G}({\omega},{\theta}^{-1}x_{21})\qquad\qquad\quad
\end{eqnarray}
where the subscript $F$ denotes an extra insertion of the operator $[B^{i},B^{j}]$ into the end of the Wilson line.
For example, an insertion of an arbitrary operator $\mathcal{O}$ into the end of the Wilson line gives
\begin{equation}
{\rho}_\mathcal{O}(x)=\int\frac{d^{2p}k}{(2\pi)^{2p}}e^{ik\cdot x}{\rm Tr}\left(\mathcal{O}e^{-ik\cdot B}\right).
\end{equation}
It is reassuring that the second term of (\ref{eq:ntlo}) is similar to the result found in \cite{ar:uvir} using
field theory; however, the first term, which is of the same order, was not mentioned there. Moreover, the physical
interpretation here in terms of dipoles is quite different.

At this point, we wish to make several comments. First of all, it is clear that (\ref{eq:ntlo}) describes the
instantaneous interaction between two points $x_1$ and $x_2$, which is consistent with the long dipole picture
depicted in FIG. \ref{fig:cyli}. In fact, all one-loop matrix diagrams, which differ only by extra operator
insertions, must have a similar double trace structure, and hence, have the physical interpretation as two-body
interactions. Secondly, note that this calculation is manifestly IR safe due to the Wilsonian cutoff on frequency
and separation. However, our approach is to be contrasted with \cite{ar:uvir}, in which an \textit{ad hoc} IR
regulator is introduced as the smallest scale in the problem. Lastly, we can identify a term in (\ref{eq:ntlo})
that leads to one-loop renormalization \cite{ls:gaug}.

The renormalization comes from a UV divergence in the planar sector,  corresponding to the constant term of $\rho$.
The integral over position then factorizes into
\begin{eqnarray}
 & & \int dt d^{2p}x_1{\rho}_{FF}(x_1,t)\int\frac{d{\omega}}{2\pi}d^{2p}x_{12}\widetilde{G}({\omega},{\theta}^{-1}x_{12})^{2}\nonumber\\
 &=&  \int dt {\rm Tr}\left({[B^{i},B^{j}]}^{2}\right) \int\frac{d{\omega}d^{2p}p}{2\pi(2\pi)^{2p}}{\widetilde{G}(\omega,p)}^{2}.
\end{eqnarray}
This quantity is easily recognized as the familiar one-loop contribution to the renormalization of the operator
${\rm Tr}{[B^{i},B^{j}]}^{2}$. Although a systematic treatment of renormalization is beyond the scope of this
work, it is clear from this example that UV dipoles in planar diagrams can lead to conventional renormalization of
the theory.

The most interesting effect, however, is the long range interaction arising from the UV finite non-planar
diagrams. We have seen that these terms come from high momentum dipoles that grow long in accordance with the
UV/IR connection. Nonetheless, the analysis has been restricted to one-loop order. In the next section, we
consider some two loop contributions, which serve to illustrate some of the general features of higher order
quantum corrections.

\section{Higher order quantum corrections}

In the analysis of the last section, we found that one-loop matrix diagrams naturally lead to double trace
operators in the effective action, which had the physical interpretation of instantaneous two-body interactions
that were mediated by long dipoles. In this section, we will study some higher order loop effects. It's not hard
to see that these diagrams will involve more traces and will, therefore, lead to instantaneous multi-body
interactions. However generically, UV divergences also appear, which can lead to strong quantum corrections.

A simple example that illustrates some of the features of higher order corrections is the analysis of the two-loop
diagrams that come from treating the quartic interactions to first order in perturbation theory
\begin{center}
\begin{minipage}[c]{2.7cm}
\includegraphics[scale=0.4]{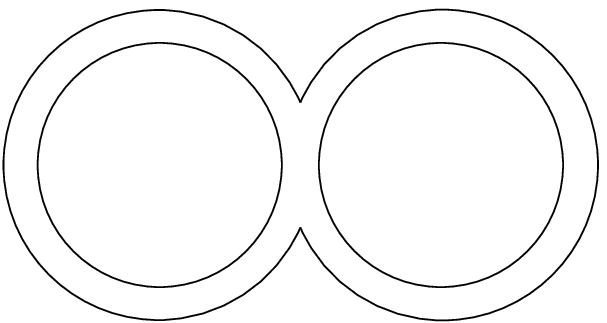}
\end{minipage}%
\begin{minipage}[c]{0.5cm}
$+$
\end{minipage}%
\begin{minipage}[c]{2.7cm}
\includegraphics[scale=0.4]{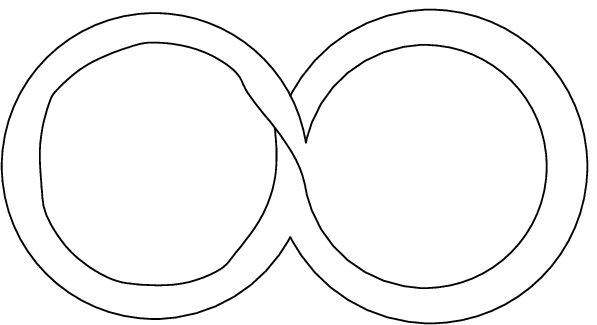}.
\end{minipage}
\end{center}
As discussed in  APPENDIX \ref{ap:two-loop}, the non-planar matrix diagram corresponds to field theory diagrams
with the two loops linked in a non-planar fashion; however, its contribution to the Wilsonian integration is
negligible. Keeping only the contribution from the planar matrix diagram, we arrive at the result derived in the
appendix. {\setlength\arraycolsep{2pt}
\begin{eqnarray}\label{eq:threebody}
 & & \int d^{2p}x_1 d^{2p}x_2 d^{2p}x_3 \rho(x_1,t)\rho(x_2,t)\rho(x_3,t)\int
 \frac{d{\omega}_1}{2\pi}\frac{d{\omega}_2}{2\pi} \nonumber\\
 & & \qquad\times \widetilde{G}({\omega}_1,{\theta}^{-1}x_{13})\widetilde{G}({\omega}_2,{\theta}^{-1}x_{23}) \nonumber\\
 & \sim & \int d^{2p}x_1 d^{2p}x_2 d^{2p}x_3 \rho(x_1,t)\rho(x_2,t)\rho(x_3,t) \nonumber\\
 & & \qquad \times \frac{1}{|x_1-x_3|}\frac{1}{|x_2-x_3|}.
\end{eqnarray}}

\begin{figure}
\psfrag{x1}{$x_1$} \psfrag{x2}{$x_2$} \psfrag{x3}{$x_3$} \psfrag{k1}{$k_1$} \psfrag{k2}{$k_2$}
\psfrag{k2-k1}{$k_3$} \psfrag{Po=(x-y)}{} \psfrag{Po=(y-z)}{} \subfigure[Loops of high momentum dipoles that
are long in the transverse direction, as discussed in this work.]{\label{quartic:subfig:a}\begin{minipage}[b]{.5\textwidth}\centering
\includegraphics[scale=0.5]{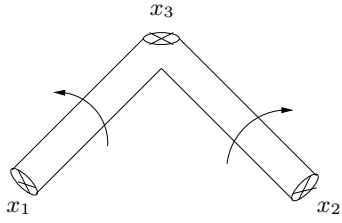}\end{minipage}}

\subfigure[Tree of low momentum dipoles that are small in the transverse
direction, as discussed in \cite{ki:inte}.]{\label{quartic:subfig:b}\begin{minipage}[b]{.5\textwidth}\centering
\includegraphics[scale=0.5]{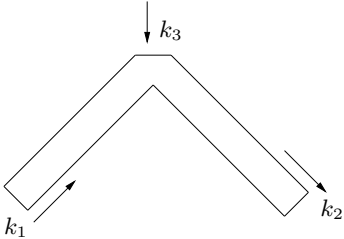}\end{minipage}}
\caption{Different dipole interpretations of the first order quartic interaction.}
\end{figure}

Using the same splitting scheme $\rho(x)={\rm tr}_{N}({\rlap{1} \hskip 1.6pt \hbox{1}})+\Delta(x)$, we can extract
the contributions from planar and non-planar field theory diagrams based on the powers of $\Delta$. For example,
the ${\Delta}^{3}$ term, which descends from purely non-planar field theory diagrams, is UV finite and corresponds
to an instantaneous three-body interaction as illustrated in FIG.~\ref{quartic:subfig:a}. At large separations,
the interaction strength falls off, so this term is consistent with a weakly coupled long distance description.

On the other hand, the terms with fewer than three $\Delta$ fields all have UV divergences arising from planar
sub-diagrams in the field theory. First of all, note that the ${\Delta}^{0}$ and ${\Delta}^{1}$ terms are
independent of the background field configuration, and therefore, do not contribute to the dynamics. We ignore
these divergent terms. However, there are two distinct divergences in the ${\Delta}^{2}$ terms which have
non-trivial consequence. The first divergence comes from the constant term in either $\rho(x_1)$ or $\rho(x_2)$.
In this case the integral factorizes into
\begin{eqnarray}
 & & \int d^{2p}x_1 d^{2p}x_3 \Delta(x_1,t)\Delta(x_3,t)\int \frac{d\omega_1}{2\pi}\widetilde{G}(\omega_1,{\theta}^{-1}x_{13})\nonumber\\
 & & \quad\times\int\frac{d\omega_2 d^{2p}p_2}{2\pi(2\pi)^{2p}}\widetilde{G}(\omega_2,p_2).
\end{eqnarray}
This quantity is a quantum correction to the leading two-body interaction (\ref{eq:non1}). The UV divergent loop
integration can be viewed as renormalizing the coupling of this operator, which is not shown explicitly. Moreover,
since the interaction strength falls with separation, this term is consistent with our perturbative analysis.
However, the other divergence coming from the constant term in $\rho(x_3)$ will lead to strong interactions
\begin{eqnarray}
 & & \int d^{2p}x_1 d^{2p}x_2 \Delta(x_1,t)\Delta(x_2,t)\int\frac{d\omega_1}{2\pi}\\
& & \quad\times\int\frac{d\omega_2 d^{2p}p_3}{2\pi(2\pi)^{2p}}\widetilde{G}(\omega_{21},p_3-{\theta}^{-1}x_{12})\widetilde{G}(\omega_2,p_3)\nonumber.
\end{eqnarray}
It is clear that this interaction grows at large separation as a power $|x_1-x_2|^{2p-2}$ for $p>1$ or as
$\log{|x_1-x_2|}$ for $p=1$. Thus, these quantum corrections lead to strongly interacting long distance behavior.

More generally, by dimensional analysis, it is clear that strong long distance behavior can only come from UV
divergences in the theory. The physical reason is that, for dipole degrees of freedom, powers of separation are
the same as powers of momentum. Thus, the cancellation of strong interactions in the matrix approach is the same
as cancelling UV divergences in the field theory.  It is, therefore, reasonable to conjecture that, given a theory
with enough supersymmetry, our perturbative analysis would exhibit weakly coupled long distance behavior, and
hence, be justified. We leave this interesting and important problem for future study.

The validity of perturbation theory aside, let us focus on the robust features of our work. We have derived a
matrix propagator for the quantum fields of noncommutative gauge theory that embodies the intrinsic dipole
structure of the quanta as well as the UV/IR relation between the transverse size of the dipoles and their center
of mass momentum. This tremendously clarified the physical effect of the quantum mechanical interactions. In
particular, we found that, quite generally, the leading IR interactions are mediated by virtual UV dipoles that
grow long in accordance with the UV/IR connection. This picture sheds new light on UV/IR mixing in noncommutative
gauge theory, reminiscent of \cite{ik:bi}. In fact, the intuition that we have developed seems very generic and
should apply to any brand of noncommutative theory. Furthermore, in this light, the non-analytic dependence of the
quantum theory on $\theta$ seems clear: in the $\theta\to 0$ limit, the quanta are no longer dipoles; therefore,
the leading long distance interactions that are present for $\theta\neq 0$ abruptly disappear, drastically
altering the IR behavior of the theory.

For some additional perspective, let us discuss the difference between our work and \cite{ki:inte}. These authors
take the star product approach to studying noncommutative scalar theory and also arrive at a dipole
interpretation.  However, they interpret the long distance interactions as an exchange of low momentum dipole
states, which is in stark contrast to the interpretation discussed here. The crucial difference is that
\cite{ki:inte} associates the $k$ variables to dipole momentum, instead of the separation $x$. In particular, the
Fourier coefficients $\tilde{\rho}(k)$ are interpreted as creating a dipole state of momentum $k$, which leads to
the scenario depicted in FIG.~\ref{quartic:subfig:b}. This picture seems to suggest a smooth $\theta\rightarrow 0$
limit, in which the transverse size of the dipoles goes to zero and the interaction becomes local. Apparently,
this interpretation is completely different from the one discussed in our work.

\begin{figure}
\psfrag{x1}{$x_1$} \psfrag{x2}{$x_2$} \psfrag{x3}{$x_3$} \psfrag{k1}{$k_1$} \psfrag{k2}{$k_2$} \psfrag{k3}{$k_3$}
\psfrag{Po=(x-y)}{} \psfrag{Po=(y-x)}{} \psfrag{Po=(z-x)}{}
\subfigure[Loops of high momentum dipoles that are
long in the transverse direction, as discussed in this work.]{\begin{minipage}[b]{.5\textwidth}\centering
\includegraphics[scale=0.5]{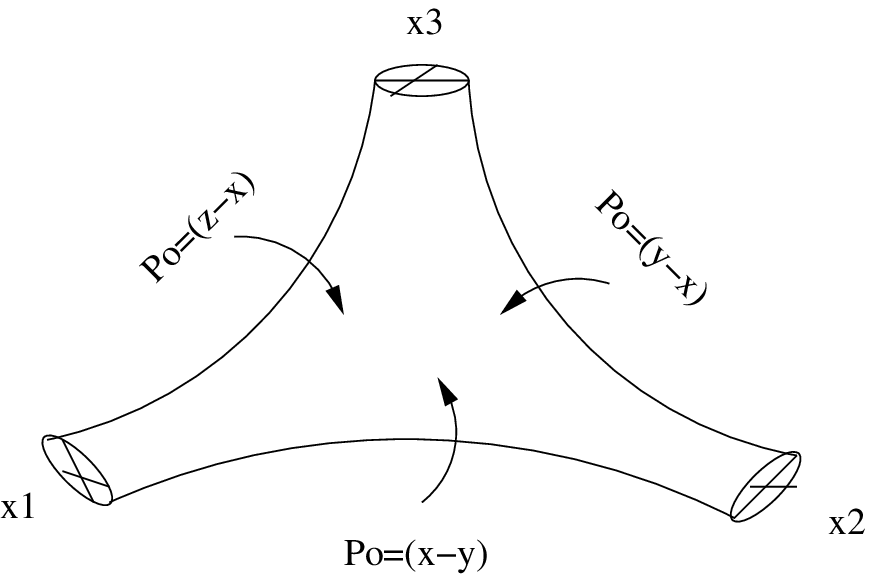}\end{minipage}}

\subfigure[Tree of low momentum dipoles that are small in the transverse
direction, as discussed in \cite{ki:inte}.]{\begin{minipage}[b]{.5\textwidth}\centering
\includegraphics[scale=0.5]{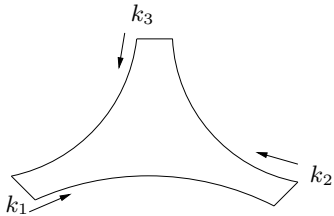}\end{minipage}}
\caption{Different dipole interpretations of the second order cubic interaction.} \label{fig:cubic}
\end{figure}

Finally, for completeness, we include the contribution from the two-loop diagram arising from the second order
treatment of the cubic interactions.
\begin{center}
\begin{minipage}[c]{1.6cm}
\includegraphics[scale=0.38]{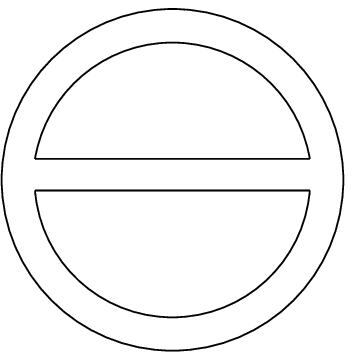}
\end{minipage}%
\begin{minipage}[c]{0.5cm}
$+$
\end{minipage}%
\begin{minipage}[c]{1.6cm}
\includegraphics[scale=0.38]{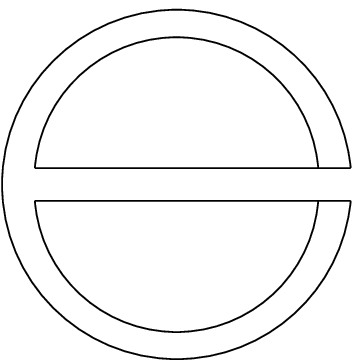}.
\end{minipage}
\end{center}
Using by now familiar techniques, we get a result proportional to
\begin{eqnarray}
\lefteqn{ \int d^{2p}x_1 d^{2p}x_2 d^{2p}x_3 \rho(x_1,t)\rho(x_2,t)\rho(x_3,t)
    \int \frac{d\omega_1}{2\pi}\frac{d\omega_2}{2\pi} }\nonumber\\
 & & \times\omega_1\widetilde{G}(\omega_1,{\theta}^{-1}x_{12})\omega_2\widetilde{G}(\omega_2,{\theta}^{-1}x_{23})
              \widetilde{G}(\omega_{12},{\theta}^{-1}x_{31}).\quad\quad
\end{eqnarray}
This process is illustrated in FIG.~\ref{fig:cubic}. However, we will not pursue the analysis of this term any
further since it is essentially identical to the analysis of (\ref{eq:threebody}).

Before closing this section we would like to emphasize the advantage of the matrix formulation for calculating
long distance interactions. The main simplification is that the noncommutative gauge invariance is manifest, and
hence, the Wilson line structure emerges automatically. Furthermore, the physical interpretation in terms of
dipole degrees of freedom is clarified tremendously. However, we have only studied the leading
order long distance behavior. Higher order terms in the derivative expansion must be retained in order to probe
the short distance structure of the theory.

\section{Discussion and outlook}

In this work, we have calculated the Wilsonian quantum effective action of noncommutative gauge theory in order to
gain some intuition for the IR dynamics of this system.  We found interaction terms suggestive of the dipole
degrees of freedom that are expected from the decoupling limit of open string theory in a strong NS-NS $B$ field.
Moreover, the leading IR interactions were mediated by long UV dipoles. In fact, this is the origin of the UV/IR
mixing: UV dipoles grow long and mediate long range interaction that dominate in the IR. This behavior  sheds new
light on the non-analytic dependence of the quantum theory on $\theta$, which  classically, is a smooth
deformation.

Perhaps the most satisfying of our results was that we constructed a representation for the matrix propagator that
embodies the intrinsic dipole structure of the elementary quanta of NCYM and the corresponding  UV/IR connection.
This approach vastly simplified calculations of the long distance interactions, as well as the physical
interpretation in terms of interacting dipoles. However, our perturbative analysis is valid only when theory is
weakly coupled at long distances. We argued that this is the case in supersymmetric theories, where UV quantum
corrections are under control.

Finally, there remain  some open questions. Most notable is that the short distance behavior is still unknown. Our
analysis is insufficient to describe the quantum effects of short dipoles, which are very sensitive to the
background field configuration. In this case, higher order terms in the derivative expansion must be retained. Of
course, we expect some short distance corrections in the form of star products in the interaction terms; however,
it would be very interesting to see some other novel effect from noncommutativity.

\begin{acknowledgments}
We thank Willy Fischler and Sonia Paban for reading this manuscript.  This work was supported by NSF grant PHY-0071512.
\end{acknowledgments}

\appendix
\section{Derivation of matrix propagator} \label{ap:propagator}

In order to have a field theory interpretation of the matrix propagator, we seek a representation of the form
\begin{equation}
\frac{1}{{\omega}^{2}-M^{2}}=\int\frac{d^{2p}k}{(2\pi)^{2p}}e^{-ik\cdot(B\otimes 1-1\otimes B)}\tilde{f}(k).
\end{equation}
The Fourier coefficients, $\tilde{f}(k)$, can be constrained by acting with ${\omega}^{2}-M^{2}$
\begin{eqnarray}
  1 &=& \int\frac{d^{2p}k}{(2\pi)^{2p}}({\omega}^{2}-M^{2})e^{-ik\cdot(B\otimes 1-1\otimes B)}\tilde{f}(k)\\
 &=& \int\frac{d^{2p}k}{(2\pi)^{2p}}\left({\omega}^{2}+\partial_k^{2}\right)e^{-ik\cdot(B\otimes 1-1\otimes B)}\tilde{f}(k)+\ldots\nonumber
\end{eqnarray}
where the $\ldots$ represent commutator terms that are necessary to resolve the ordering of the noncommuting
matrices, $B^{i}\otimes {\rlap{1} \hskip 1.6pt \hbox{1}}-{\rlap{1} \hskip 1.6pt \hbox{1}}\otimes B^{i}$. It is
easy to see that the commutator corrections are negligible if
\begin{equation}
B^{i}\gg [k\cdot B,B^{i}], \left[k\cdot B,[k\cdot B,B^{i}]\right],\ldots
\end{equation}
Using the expression for the background field,
$B^{i}={\hat{x}}^{i}\otimes {\rlap{1} \hskip 1.6pt \hbox{1}}_{N\times N}+{\theta}^{ij}A_j(\hat{x})$,
we see that $[k\cdot B,\quad]=k\cdot\theta\cdot D$ where $D_i$ is the gauge covariant derivative. Therefore
the commutators are small if $\theta\cdot k\ll L$, $L$ being the length scale set by the curvature of the background.
Note that $L<\sqrt{\theta}$ because the NCYM dual description includes a constant background field strength of
magnitude ${\theta}^{-1}$.

Assuming that the commutators are negligible, we may keep only the leading term in above equation. Then, upon an
integration by parts, the condition on $\tilde{f}$ becomes
\begin{equation}
\left({\omega}^{2}+\partial_k^{2}\right)\tilde{f}(k)=(2\pi)^{2p}{\delta}^{2p}(k).
\end{equation}
From this equation, we arrive at the integral expression
\begin{equation}
\tilde{f}(k)=\int d^{2p}x\frac{e^{ik\cdot x}}{{\omega}^{2}-x^{2}}.
\end{equation}
Note that the consistency condition $\theta\cdot k\ll L$ is equivalent to $x\gg {\theta}/{L}$. We, therefore,
apply a cutoff $x>\theta\Lambda\gg {\theta}/{L}$. Thus, up to commutator terms that are suppressed by factors
of $(L\Lambda)^{-1}\ll 1$, we obtain the desired representation
{\setlength\arraycolsep{2pt}
\begin{eqnarray}
& & \frac{1}{{\omega}^{2}-M^{2}} \nonumber\\
& = & \int\frac{d^{2p}k}{(2\pi)^{2p}}e^{-ik\cdot(B\otimes 1-1\otimes B)}\int_{\theta\Lambda} d^{2p}x
\frac{e^{ik\cdot x}}{{\omega}^{2}-x^{2}}.
\end{eqnarray}}

\section{Next to leading order one-loop diagram} \label{ap:one-loop}

The qualitative features of the next to leading order one-loop behavior is contained in FIG.~\ref{fig:subfig:a}
with two extra insertions of the field strength. A straight forward perturbative treatment of the field strength
term in $L_2$ gives
\begin{widetext}
\begin{eqnarray}
& & \int dt_1 d^{2p}x_1 dt_2 d^{2p}x_2\int\frac{d{\omega}_1 d^{2p}k_1}{2\pi(2\pi)^{2p}}\frac{d{\omega}_2
d^{2p}k_2}{2\pi(2\pi)^{2p}}e^{-i{\omega}_1(t_1-t_2)+ik_1\cdot x_1}e^{-i{\omega}_2(t_2-t_1)+ik_2\cdot x_2}
\times \widetilde{G}({\omega}_1,{\theta}^{-1}x_1)\widetilde{G}({\omega}_2,{\theta}^{-1}x_2) \nonumber\\
& & \qquad\qquad\qquad \times \left[{\rm Tr}\left([B^{i},B^{j}](t_1)e^{ik_1\cdot B(t_1)}[B^{i},B^{j}](t_2)e^{-ik_2\cdot B(t_2)}\right){\rm Tr}\left(e^{-ik_1\cdot B(t_1)}e^{ik_2\cdot B(t_2)}\right)\right. \nonumber\\
& & \qquad\qquad\qquad \left.{}-{\rm Tr}\left([B^{i},B^{j}](t_1)e^{ik_1\cdot B(t_1)}e^{-ik_2\cdot B(t_2)}\right){\rm Tr}\left([B^{i},B^{j}](t_2)e^{-ik_1\cdot B(t_1)}e^{ik_2\cdot B(t_2)}\right)\right].
\end{eqnarray}
\end{widetext}
Since we only integrate out virtual states with high energy and momentum, time derivatives of the background as
well as higher commutator terms are further suppressed. Therefore, to lowest order, we obtain
\begin{widetext}
\begin{eqnarray}
& & \int dt d^{2p}x_1 d^{2p}x_2\int\frac{d\omega}{2\pi}\frac{ d^{2p}k_1}{(2\pi)^{2p}}\frac{d^{2p}k_2}{(2\pi)^{2p}}e^{ik_1\cdot x_1}e^{ik_2\cdot x_2}
\widetilde{G}({\omega},{\theta}^{-1}x_1)\widetilde{G}({\omega},{\theta}^{-1}x_2)
\left[{\rm Tr}\left([B^{i},B^{j}](t)[B^{i},B^{j}](t)e^{i(k_1-k_2)\cdot B(t)}\right)\right. \nonumber\\
& & \qquad\qquad\quad \left. \times {\rm Tr}\left(e^{-i(k_1-k_2)\cdot B(t)}\right)
-{\rm Tr}\left([B^{i},B^{j}](t)e^{i(k_1-k_2)\cdot B(t)}\right){\rm Tr}\left([B^{i},B^{j}](t)e^{-i(k_1-k_2)\cdot B(t)}\right)\right].
\end{eqnarray}
\end{widetext}
after performing one integral over $\omega$ and one integral over $t$. Upon Fourier transforming to position space, we are finally left with (\ref{eq:ntlo}).

\section{Two-loop example} \label{ap:two-loop}

A straight forward evaluation of the quartic two-loop diagrams turns out to be proportional to
\begin{eqnarray}\label{eq:twoloop}
& & \int dt d^{2p}x_1 d^{2p}x_2\int\frac{d{\omega}_1 d^{2p}k_1}{2\pi(2\pi)^{2p}}\frac{d{\omega}_2
d^{2p}k_2}{2\pi(2\pi)^{2p}}e^{ik_1\cdot x_1}e^{ik_2\cdot x_2} \nonumber\\
& & \quad \times \widetilde{G}({\omega}_1,{\theta}^{-1}x_1)\widetilde{G}({\omega}_2,{\theta}^{-1}x_2) \nonumber\\
& & \times \left[{\rm Tr}\left(e^{ik_1\cdot B(t)}\right){\rm Tr}\left(e^{-ik_2\cdot B(t)}\right){\rm
Tr}\left(e^{-ik_1\cdot B(t)}e^{ik_2\cdot B(t)}\right)\right. \nonumber\\
& & \left.{}-{\rm Tr}\left(e^{ik_1\cdot B(t)}e^{ik_2\cdot B(t)}e^{-ik_1\cdot B(t)}e^{-ik_2\cdot
B(t)}\right)\right].
\end{eqnarray}
As indicated by the double line diagrams, the triple trace term comes from the planar matrix diagram, and the
single trace term comes from the non-planar matrix diagram.

To understand the meaning of the non-planar matrix diagram, consider the corresponding field theory vacuum
diagram. As previously discussed, the vacuum diagrams are obtained by setting $A_i=0$, in which case the background field $B^{i}=\hat{x}^{i}\otimes {\rlap{1} \hskip 1.6pt \hbox{1}}_{N\times N}$. Substituting this background field into the single trace term above, we find a result proportional to
\begin{eqnarray}\label{eq:vac}
& & \int dt d^{2p}x_1 d^{2p}x_2\int\frac{d{\omega}_1 d^{2p}k_1}{2\pi(2\pi)^{2p}}\frac{d{\omega}_2
d^{2p}k_2}{2\pi(2\pi)^{2p}}e^{ik_1\cdot x_1}e^{ik_2\cdot x_2} \nonumber\\
& & \quad \times \widetilde{G}({\omega}_1,{\theta}^{-1}x_1)\widetilde{G}({\omega}_2,{\theta}^{-1}x_2)
      e^{ik_1\cdot\theta\cdot k_2}{\rm Tr}({\rlap{1} \hskip 1.6pt \hbox{1}}).
\end{eqnarray}
Note that we have kept the higher order commutators, which lead to the phase factor $\exp(ik_1\cdot\theta\cdot k_2)$, only to illustrate the qualitative nature of this
contribution. Upon performing the $k$ integrals and using the identity ${\rm Tr}({\rlap{1} \hskip 1.6pt
\hbox{1}})=\int d^{2p}x_3 {\rm tr}_{N}({\rlap{1} \hskip 1.6pt \hbox{1}})$, we get a term proportional to
\begin{eqnarray}\label{eq:exp}
& & \int dt d^{2p}x_3 \frac{d{\omega}_1 d^{2p}p_1}{2\pi(2\pi)^{2p}}\frac{d{\omega}_2 d^{2p}p_2}{2\pi(2\pi)^{2p}}e^{ip_1\cdot\theta\cdot p_2} \nonumber\\
& & \qquad\quad \times\widetilde{G}(\omega_1,p_1)\widetilde{G}(\omega_2,p_2),
\end{eqnarray}
Note that we have changed variables of integration to $p={\theta}^{-1}\cdot x$. This term is easily recognized as
the contribution to the effective action from the two loop non-planar field theory vacuum diagram below
\begin{center}
\includegraphics[scale=0.5]{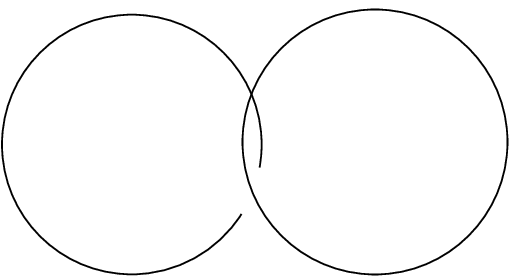}.
\end{center}
Thus, the non-planar matrix diagram corresponds to field theory diagrams with the loops linked in a non-planar
fashion. Furthermore, it is easy to see from the single trace term in (\ref{eq:twoloop}) that the effect of the
background gauge field insertions is simply to include a series of higher dimensional field theory operators in
the integral over $x_3$, and for dimensional reasons, more powers of $p$ in the denominator. The integration over
$\omega,p$ remains decoupled from the integral over $x_3$, though.

However, a closer look at the non-planar matrix diagram reveals that it is negligible in the domain of Wilsonian
integration. In fact, the contribution is non-perturbative in our expansion parameter $(L\Lambda)^{-1}$. Essentially, the reason is that the phase factor $\exp(ip_1\cdot\theta\cdot p_2)$ is rapidly oscillating for dipole momenta $p$ above the Wilsonian cutoff, which gives rise to exponential suppression. To see this, consider the integration over $\omega,p$ in (\ref{eq:exp})
\begin{eqnarray}
 & & \int_{\Lambda}\frac{d{\omega}_1 d^{2p}p_1}{2\pi(2\pi)^{2p}}\frac{d{\omega}_2 d^{2p}p_2}{2\pi(2\pi)^{2p}} e^{ip_1\cdot\theta\cdot p_2}\widetilde{G}(\omega_1,p_1)\widetilde{G}(\omega_2,p_2)\nonumber\\
 &\sim & \int_{\Lambda} \frac{d^{2p}p_1}{|\theta p_1|}\frac{d^{2p}p_2}{|\theta p_2|} e^{ip_1\cdot\theta\cdot p_2}
               \sim {\Lambda}^{4p-2}e^{-\theta{\Lambda}^{2}}
\end{eqnarray}
where the final integration can be performed with Schwinger parameters in the stationary phase approximation. Note that $L\Lambda<\theta\Lambda$, as previously discussed. Clearly, the exponential suppression coming from the phase factor $\exp(p_1\cdot\theta\cdot p_2)$ is universal, while
the power of $\Lambda$ comes from the powers of $p$ in the integral. Thus, in the domain of Wilsonian integration,
the contribution from the nonplanar matrix diagram is exponentially suppressed, and therefore, utterly negligible.
In fact, the same argument implies that we can neglect all nonplanar matrix diagrams relative to the planar ones.

While nonplanar matrix diagrams do not contribute to the Wilsonian integration, they do seem to involve
non-trivial short distance effects. As is well known from field theory, (\ref{eq:exp}) is IR divergent without a
cutoff \cite{ms:nonp}. Furthermore, it seems natural to attribute this to short distance effects, since the IR
dipoles are small in spatial extent and only a single trace appears. Of course, it would be interesting to better
understand the quantum effects of low momentum dipoles, but this is simply beyond the validity of our
approximations. However, the matrix approach does seem to clarify the role of this IR singularity to the extent
that it is not a long distance effect. Actually, \cite{ik:bi} seems to suggest that the quantum effects of low
momentum states in noncommutative theories should be the same as that in ordinary theories.

Proceeding with the calculation, we keep only the triple trace term. To leading order, we neglect the commutators
in (\ref{eq:twoloop}), in which case we are left with
\begin{eqnarray}
\lefteqn{ \int dt d^{2p}x_1 d^{2p}x_2 d^{2p}x_3 \rho(x_1,t)\rho(x_2,t)\rho(x_3,t)\int
 \frac{d{\omega}_1}{2\pi}\frac{d{\omega}_2}{2\pi} } \nonumber\\
 & & \qquad\qquad \times \widetilde{G}({\omega}_1,{\theta}^{-1}x_{13})\widetilde{G}({\omega}_2,{\theta}^{-1}x_{23})
     \qquad\qquad\quad
\end{eqnarray}
after a Fourier transformation to position space.

\end{document}